\documentclass[twocolumn,showpacs,superscriptaddress]{revtex4}%
\usepackage{graphicx}%
\usepackage{amsmath}
\usepackage{lmodern}     % set math font to Latin modern math
\usepackage[T1]{fontenc}
\usepackage{epstopdf}
\DeclareSymbolFont{myletters}{OML}{ztmcm}{m}{it}
\DeclareMathSymbol{\uplambda}{\mathord}{myletters}{"15}
\usepackage{amsfonts}
\usepackage{amsmath}
\usepackage{amssymb}
\usepackage{gensymb}
\usepackage{xcolor}
\providecommand{\U}[1]{\protect\rule{.1in}{.1in}}

\begin{document}
\title{Superconductivity in two-dimensional phosphorus carbide ($\beta_{0}$-PC)}
\author{Bao-Tian Wang}
\thanks{E-mail: wangbt@ihep.ac.cn }
 \affiliation{Institute of High Energy Physics, Chinese Academy of
Sciences (CAS), Beijing 100049, China} \affiliation{Dongguan
Institute of Neutron Science (DINS), Dongguan 523808, China}
\author{Peng-Fei Liu}
\thanks{E-mail: pfliu@ihep.ac.cn }
 \affiliation{Institute of High Energy Physics, Chinese Academy of
Sciences (CAS), Beijing 100049, China} \affiliation{Dongguan
Institute of Neutron Science (DINS), Dongguan 523808, China}
\author{Tao Bo}
 \affiliation{Institute of High Energy Physics, Chinese Academy of
Sciences (CAS), Beijing 100049, China} \affiliation{Dongguan
Institute of Neutron Science (DINS), Dongguan 523808, China}
\author{Wen Yin}
\affiliation{Institute of High Energy Physics, Chinese Academy of
Sciences (CAS), Beijing 100049, China} \affiliation{Dongguan
Institute of Neutron Science (DINS), Dongguan 523808, China}
\author{Fangwei Wang}
\affiliation{Dongguan Institute of Neutron Science (DINS), Dongguan
523808, China} \affiliation{Beijing National Laboratory for
Condensed Matter Physics, Institute of Physics, Chinese Academy of
Sciences (CAS), Beijing 100080, China}

\begin{abstract}
{Two-dimensional (2D) boron has been predicted to show
superconductivity. However, intrinsic 2D carbon and phosphorus have
not been reported to be superconductors, which, inspires us to seek
superconductivity in their mixture. Here we perform first-principles
calculations of the electronic structure, phonon dispersion, and
electron-phonon coupling of the metallic phosphorus carbide
monolayer, the $\beta_{0}$-PC. Results show that it is an intrinsic
phonon-mediated superconductor, with estimated superconducting
temperature $T_{c}$ to be $\sim$13 K. The main contribution to the
electron-phonon coupling is from the out-of-plane vibrations of
phosphorus. A Kohn anomaly on the first acoustic branch is observed.
The superconducting related physical quantities is found tunable by
applying strain or carrier doping.}

\end{abstract}

\maketitle

Owing to the developments of the atomic techniques, like molecular
beam epitaxy \cite{MBE}, atomic layer deposition \cite{ALD}, pulsed
laser deposition \cite{PLD}, magnetron sputtering \cite{Haberkorn},
etc., many two-dimensional (2D) or layered superconductors have been
synthesized successfully \cite{Uchihashi,Brun}. The
superconductivity can be remarkably robust in the 2D limit with
respective to the corresponding parental bulk materials
\cite{GeJF,Clark}. The fascinating phenomena of the competition with
charge density waves \cite{XiNbSe2}, the Kohn anomaly \cite{Yan},
and the strong spin-orbit coupling \cite{Nam} for superconductivity
in 2D systems have attracted many attentions.

Since the well-known discovery of graphene by mechanical exfoliation
\cite{Novoselov1,Novoselov2}, various 2D-monolayer carbon
\cite{Graphdiyne,2D-Carbon} and phosphorus \cite{BlackP,BlueP} have
been successfully obtained. Both graphene and phosphorene can be
manipulated to be superconductors through carrier doping
\cite{Savini,Si,MarginePRB,Shao}, strain \cite{Si,Ge}, and/or
metal-decorating/intercalating
\cite{Ludbrook,MargineSR,Zheng,Huang,ZhangR}. However, till now, no
superconducting properties have been experimentally observed or
theoretically predicted for these famous 2D materials in their
intrinsic forms. This fact stimulates our interests to seek
superconducting state in their compounds, \emph{i.e.} the 2D
phosphorus carbide (PC).

Recently, various allotropes of 2D-monolayer PC were theoretically
predicted \cite{WangPC,Guan} to be stable through the particle-swarm
optimization method \cite{CALYPSO} and the density functional theory
(DFT) \cite{VASP2}. Later on, few-layer 2D black PC has been
synthesized successfully via a novel carbon doping technique
\cite{Tan}. Among all reported 2D PC, only the graphene-like
$\beta_{0}$-PC exhibits metallic character \cite{Guan,Rajbanshi} and
has been testified carefully to be stable in a DFT level
\cite{Rajbanshi}. Therefore, it is necessary to verify whether this
newly reported phase exhibits superconductivity or not.

The calculations are performed at the DFT level, employing the local
density approximation and norm-conserving pseudopotentials
\cite{Troullier,Fuchs} as implemented in the QUANTUM-ESPRESSO (QE)
package \cite{Giannozzi}. The plane-waves kinetic-energy cutoff is
set as 100 Ry and the structural optimization is performed until the
forces on atoms are less than 10 meV/\AA. Monolayer PC is simulated
with a vacuum thickness of 20 \AA, which is enough to decouple the
adjacent layers. An unshifted Brillouin-zone (BZ) \textbf{k}-point
mesh of 16$\times$6 and a Hermitian-Gaussian smearing method are
adopted for the electronic charge density calculations. The phonon
modes are computed within density-functional perturbation theory
\cite{Baroni} on a 8$\times$3 \textbf{q} mesh.

\begin{figure}
\begin{center}
\includegraphics[width=1\linewidth]{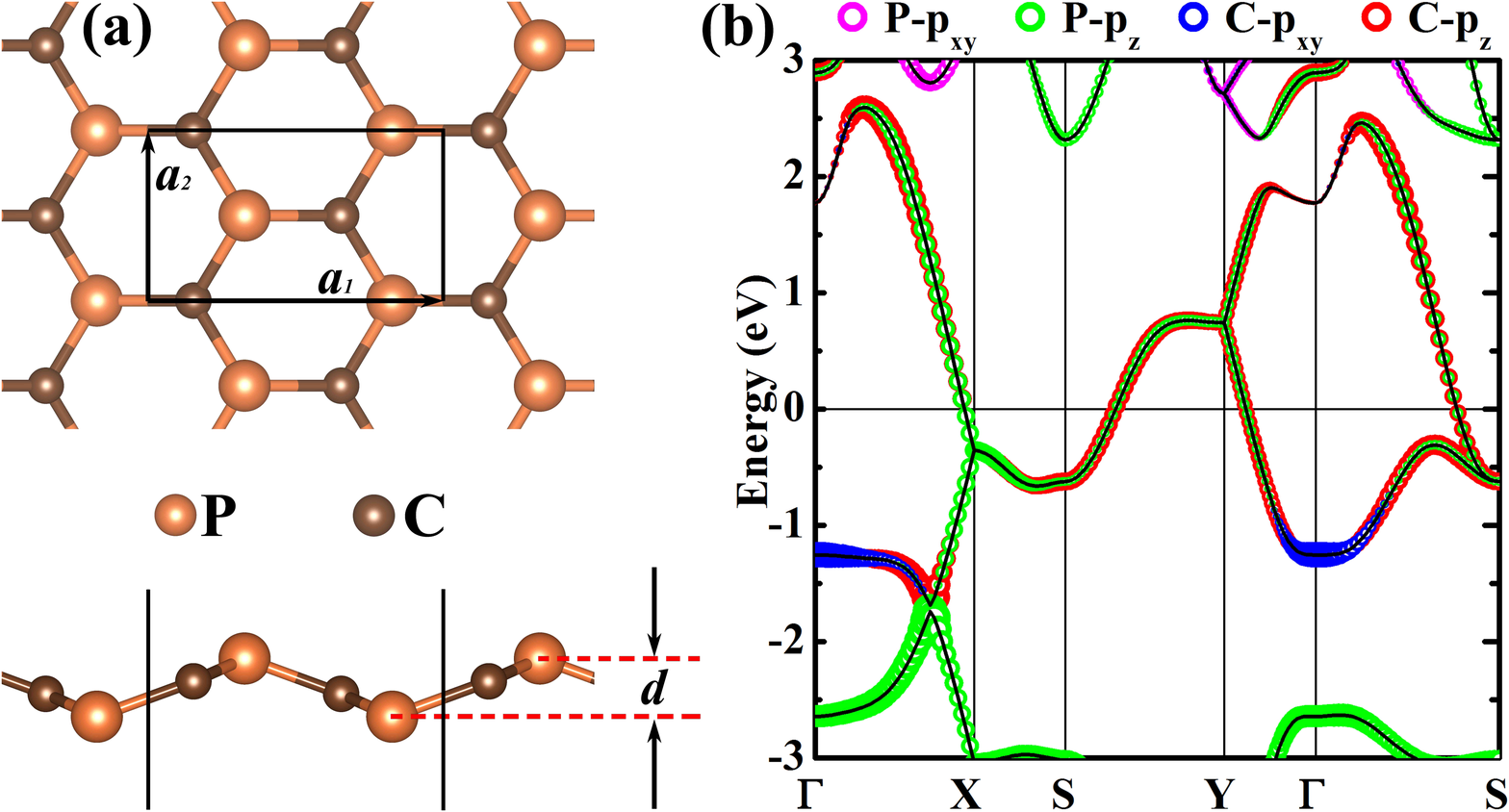}
\end{center}
\caption{Crystal and band structures of $\beta_{0}$-PC. (a)
Schematic of the crystal structure with the unit cell indicated by
the black rectangle. $a_{1}$ and $a_{2}$ are the lattice constants
while $d$ stands for the buckling thickness. (b) Orbital-resolved
band structure with the contributions of $p_{xy}$ and $p_{z}$
orbitals of
P (C) atoms being indicated by magenta and green (blue and red) hollow circles, respectively. The Fermi energy level is set at zero.}%
\label{fig1}%
\end{figure}

The optimized phosphorus carbide ($\beta_{0}$-PC) monolayer
crystallizes in the polar orthorhombic structure with space group
$Pmn21$ (no. 31), showing the $C_{2v}$ symmetry. Similar to the
planar honeycomb lattice of graphene \cite{Novoselov1,Novoselov2},
all the atoms in the unit cell of $\beta_{0}$-PC are 3-fold
coordinated, which endows the structure with topologically repeated
hexagonal configuration. According to the chemical octet rule
\cite{Langmuir}, C atoms usually feature $p$-election conjugated
bonds via $sp^{2}$-hybridization in the planar geometry
(\emph{e.g.}, graphene \cite{Novoselov1,Novoselov2}, graphenylene
\cite{Song,Yu}, and phagraphene \cite{WangPhaG}), whereas P atoms
adopt $sp^{3}$-hybridization with a lone electron pair forming the
puckered configuration of phosphorene \cite{ZhuP,GuanP}. As shown in
Figure 1a, the alternating manner of P and C atoms renders
$\beta_{0}$-PC monolayer a buckled structure, which mainly stems
from the competition between the favoured planar
$sp^{2}$-hybridization of C atoms and nonplanar
$sp^{3}$-hybridization of P atoms. $\beta_{0}$-PC monolayer shows an
anisotropic rectangular structure with lattice constants
$a_{1}$=4.99 \AA\ and $a_{2}$=2.87 \AA\
($a_{1}$/$a_{2}$$\approx$1.74), respectively, which are consistent
with reported results of $a_{1}$= 5.02 \AA\ and $a_{2}$= 2.95 \AA\
($a_{1}$/$a_{2}$$\approx$1.70) \cite{Rajbanshi}. The buckling
thickness of the $\beta_{0}$-PC monolayer, simply measured by the
distance between the top and bottom atomic layers, is $d$=0.98 \AA\
being slightly smaller than that (1.24 \AA) in the honeycomb lattice
of semiconducting blue phosphorene \cite{Ozcelik}. Our QE
calculations reveals two distinct types of P-C bond lengths varying
from 1.72 \AA\ to 1.74 \AA\ for $\beta_{0}$-PC monolayer, which are
in analogy with the reported values (1.75 \AA\ and 1.77 \AA)
\cite{Rajbanshi} obtained by VASP \cite{VASP1,VASP2} and different
from one P-C bond length of 1.78 \AA\ \cite{Guan} using SIESTA
\cite{SIESTA}. The P-C bond lengths in $\beta_{0}$-PC monolayer are
between the P-C single bond length (1.83 \AA) and the P-C double
bond length (1.67 \AA) \cite{Rajbanshi,Daly}, indicating some
greater $p$-character $sp^{2}$-hybridization for C atoms and some
less $p$-character $sp^{3}$-hybridization for P atoms, which helps
stabilize the anisotropic buckled rectangular lattice
\cite{Rajbanshi}.

The orbital-resolved band structure of $\beta_{0}$-PC is shown in
Figure 1b. $\beta_{0}$-PC monolayer exhibits intrinsic metallic
features with bands crossing the Fermi level, which is in good
agreement with the previous theoretical calculations
\cite{Rajbanshi,Guan}. In the vicinity of Fermi level, the valence
and conduction bands mostly originate from the hybridization of
P-$p_{z}$ and C-$p_{z}$ orbitals according to the partial band
projections. Thus, they come into being a high density of 2D
$\pi$-electron gas at the Fermi level via unconventional $\pi$-$\pi$
interactions which delocalize the lone pair electrons of P atoms and
bring the imperfect $sp^{2}$- and $sp^{3}$-hybridized states for
$\beta_{0}$-PC monolayer. Apparently, such characteristic gives rise
to metallic behavior for $\beta_{0}$-PC monolayer, and accordingly,
favours a superconducting sheet.

\begin{figure}[ptb]
\begin{center}
\includegraphics[width=1\linewidth]{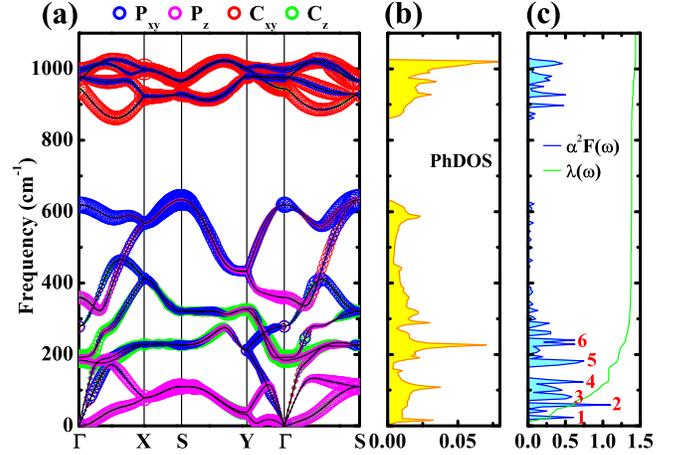}
\end{center}
\caption{(a) Phonon frequency dispersion of $\beta_{0}$-PC weighted
by the motion modes of P and C atoms. The blue, magenta, red, and
green hollow circles indicate P horizontal, P vertical, C
horizontal, and C vertical modes, respectively. (b) PhDOS and (c)
Eliashberg spectral function $\alpha^{2}$F($\omega$) and cumulative
frequency-dependent of EPC $\uplambda$($\omega$).
The peaks of $\alpha^{2}$F($\omega$) are numbered by vibration frequency from 1 to 6.}%
\label{fig2}%
\end{figure}

We now focus on the vibration properties and the electron-phonon
coupling (EPC) in $\beta_{0}$-PC. Figure 2a shows the phonon
dispersion over the whole BZ. The absence of the imaginary modes
clearly indicates that the $\beta_{0}$-PC is dynamically stable. Our
results agree well with one recent calculation \cite{Rajbanshi} by
Rajbanshi \emph{et al.} who use the finite displacement method in
obtaining phonon spectra. From the decomposition of the phonon
spectrum with respect to C and P atomic vibrations, as indicated in
Figure 2a, we find that the main contribution to the acoustic
branches below 150 cm$^{-1}$ is the phosphorus out-of-plane P$_{z}$
vibrations. The interaction between P and C atoms contributes to the
intermediate-frequency region from 150 to 634 cm$^{-1}$. The
in-plane modes of C atoms occupy the high frequencies above 860
cm$^{-1}$. Similar to Li-decorated monolayer graphene \cite{Zheng}
and 2D Cu-benzenehexathial (Cu-BHT) \cite{Zhang}, here, the vertical
vibrations of C atoms are also lower than their horizontal modes.

The phonon density of state (PhDOS), the Eliashberg electron-phonon
spectral function $\alpha^{2}$F($\omega$), and the cumulative
frequency-dependent of EPC $\uplambda$($\omega$) are displayed in
Figures 2b and 2c. Here, the $\alpha ^{2}F(\omega)$ and the
$\uplambda$($\omega$) are calculated by
\begin{align}
\alpha^{2}F(\omega)=\frac{1}{2\pi\!N(E_{\mathrm{{F}}})}\sum_{\textbf{\emph{q}}\nu}\frac
{\gamma_{\textbf{\emph{q}}\nu}}{\omega_{\textbf{\emph{q}}\nu}}\delta(\omega-\omega_{\textbf{\emph{q}}\nu})
\end{align}
and
\begin{align}
\uplambda(\omega)=2\int_{0}^{\omega}\frac{\alpha^{2}\emph{F}(\omega)}{\omega}d\omega,
\end{align}
where $N$(\emph{E}$_{\mathrm{{F}}}$) is the electronic density of
state at the Fermi level and $\gamma_{\textbf{\emph{q}}\nu}$ is the
phonon linewidth. We find that the low-frequency phonons, mainly
associated with the out-of-plane P$_{z}$ modes, are key to achieving
a high EPC in $\beta_{0}$-PC as they account for 1.08 (73\%) of the
total EPC ($\uplambda$=1.48). As shown in Figure 2c, the 1st to 4th
peaks of the $\alpha^{2}$F($\omega$) are responsible for this part.
The 5th and the 6th peaks hold the main feature of the out-of-plane
C$_{z}$ modes and associate with a EPC strength of 0.23 (16\%).
Similar to graphene \cite{Si} and 2D Cu-BHT \cite{Zhang}, the EPC
induced by high-frequency phonons is almost negligible.
Specifically, the high-frequency phonons, mainly the C$_{xy}$ modes,
only contribute 0.06 (4\%) of the total EPC. Overall, our calculated
EPC value of 1.48 clearly makes the 2D $\beta_{0}$-PC an
intermediate to strong conventional superconductor.

\begin{figure}
\begin{center}
\includegraphics[width=1\linewidth]{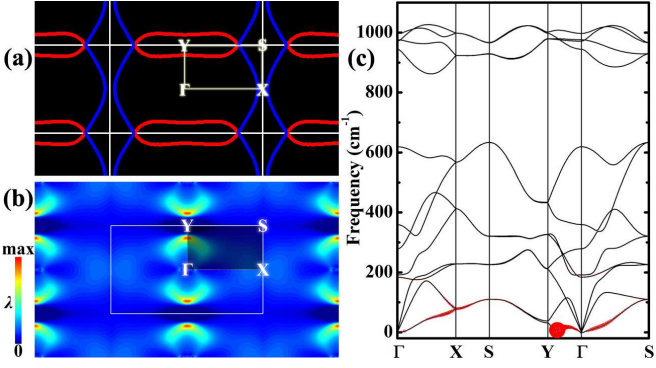}
\end{center}
\caption{(a) Fermi surface in BZ with hole and electron bands being
indicated by red and green contour lines, respectively. (b) The
distribution of \textbf{q}-resolved EPC
$\uplambda_{\textbf{\emph{q}}}$ for the first transverse acoustic
branch in BZ. (c) Phonon spectrum with the size of red dots being
drawn proportional to the magnitude
of the EPC $\uplambda_{\textbf{\emph{q}}\nu}$.}%
\label{fig3}%
\end{figure}

The Fermi surface contour of $\beta_{0}$-PC is shown in Figure 3a,
where one can see one electron pocket centered at the Y point. Away
from the $\Gamma$ point and along S-X-S, a hole arc is emerged.
Along Y-S and near the S point, the electron pocket crosses with the
hole arc. Given the electronic states of P and C atoms at the Fermi
level, the out-of-plane phonon vibrations couple strongly with the
2D $\pi$-electron hybridization of P-$p_{z}$ and C-$p_{z}$. As a
result, the distribution of the $\uplambda_{\textbf{\emph{q}}}$
($\uplambda_{\textbf{\emph{q}}\nu}$=$\frac{\gamma_{\textbf{\emph{q}}\nu}}{\pi\!hN(E_{\mathrm{{F}}})\omega_{\textbf{\emph{q}}\nu}^{2}}$)
for the first transverse acoustic (TA) branch in BZ should overlap
with the electronic Fermi surface. In Figure 3b, an arc-type
distribution of $\uplambda_{\textbf{\emph{q}}}$ is observed and is
believed to strongly couple with the electron pocket near the Y
point. Combined Figure 3b with Figure 3c, one can find that the
largest value of the EPC appears at \textbf{q}$_{1}$=(0,0.36) point
along $\Gamma$-Y on the first TA branch. This Kohn anomaly
\cite{Kohn,Caruso} or softening of the phonon mode yields
significant coupling between electrons and acoustic phonons, with a
very low frequency of $\sim$8 cm$^{-1}$.

\begin{table}[ptb]
\caption{The superconductive parameters of $\mu^{*}$,
$N$(\emph{E}$_{\mathrm{{F}}}$) (in unit of states/spin/Ry/cell),
$\omega$$\rm{_{log}}$ (in K), $\uplambda$, and $T_{c}$ (in K) for
some intrinsic 2D-monolayer-phonon-mediated superconductors.}%
\begin{ruledtabular}
\begin{tabular}{lccccccccccccccccccc}
Compounds&$\mu^{*}$&$N$(\emph{E}$_{\mathrm{{F}}}$)&$\omega$$\rm{_{log}}$&$\uplambda$&$T_{c}$&Refs.\\
\hline
B$_{2}$C&0.1&&314.8&0.92&19.2&\cite{Dai}\\
B$_{\triangle}/$B$_{\Box}$/B$_{\diamondsuit}$&0.1&&&1.1/0.8/0.6&21/16/12&\cite{Penev}\\
B ($\alpha$ sheet)&0.05&5.85&262.2&0.52&6.7&\cite{ZhaoB}\\
B ($\beta_{12}$)&0.1-0.15&8.12&425&0.69&14&\cite{Cheng}\\
TiSi$_{4}$&0.1&&&0.59&5.8&\cite{Wu}\\
Li$_{2}$B$_{7}$&0.12&&462.8&0.56&6.2&\cite{WuLiB}\\
Mo$_{2}$C&0.1&&&0.63&5.9&\cite{ZhangMo2C}\\
Cu-BHT&0.1&&51.8&1.16&4.43&\cite{Zhang}\\
$\beta_{0}$-PC&0.1&7.27&118.0&1.48&13.35&This work\\
\end{tabular}
\label{Tc}
\end{ruledtabular}
\end{table}

Using our calculated Eliashberg spectral function
$\alpha^{2}$F($\omega$) and $\uplambda$, we calculate the
logarithmic average frequency defined as $\omega$$\rm{_{log}}$=exp$\left[  \frac{2}{\uplambda}\int_{0}^{\infty}%
\frac{\emph{d}\omega}{\omega}\alpha^{2}\emph{F}(\omega)\rm{_{log}}\omega\right]$
to be 118.0 K. Using a typical value of the effective screened
Coulomb repulsion constant $\mu^{*}$=0.1, the superconducting
transition temperature $T_{c}$ can be estimated, according to the
Bardeen-Cooper-Schrieffer (BCS) theory \cite{Bardeen}, by the
Allen-Dynes modified McMillan equation \cite{Allen,McMillan}
\begin{align}
T_{c}=\frac{\omega\mathrm{{_{log}}}}{1.2}\mathrm{{exp}\left[  -\frac
{1.04(1+\uplambda)}{\uplambda-\mu^{*}(1+0.62\uplambda)}\right].}%
\end{align}
Our calculated value of $T_{c}$ is 13.35 K, comparable with that
predicted in 2D Boron \cite{Penev,Cheng}.

Actually, superconductivity has little been predicted for intrinsic
2D-monolayer systems. In Table I, we list the superconductive
parameters of $\mu^{*}$, $N$(\emph{E}$_{\mathrm{{F}}}$),
$\omega$$\rm{_{log}}$, $\uplambda$, and $T_{c}$ for some typical 2D
phonon-mediated superconductors
\cite{Dai,Penev,ZhaoB,Cheng,Wu,WuLiB,ZhangMo2C,Zhang}. All these
systems have been predicted to appear superconductivity without
external conditions of high pressure, strain, carrier doping, metal
decorations/intercalations, and/or functional groups. It is clear
that the $T_{c}$ of $\beta_{0}$-PC is larger than that of 2D boron
($\alpha$ sheet) \cite{ZhaoB}, TiSi$_{4}$ \cite{Wu}, Li$_{2}$B$_{7}$
\cite{WuLiB}, Mo$_{2}$C \cite{ZhangMo2C}, and Cu-BHT \cite{Zhang}
while smaller than that of B$_{2}$C \cite{Dai} and B$_{\triangle}$
\cite{Penev}. Among these systems, $\beta_{0}$-PC exhibits the
largest value of $\uplambda$ but holds a relatively small value of
$\omega$$\rm{_{log}}$. The vertical vibrations of P atoms in
$\beta_{0}$-PC play similar role in EPC as that of S atoms in Cu-BHT
\cite{Zhang}.

\begin{figure}
\begin{center}
\includegraphics[width=1\linewidth]{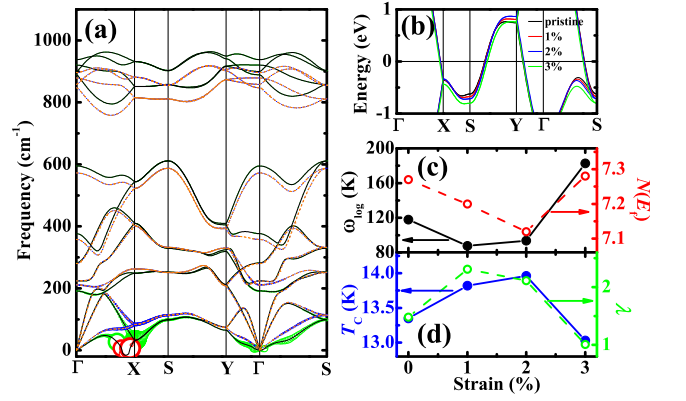}
\end{center}
\caption{Superconducting related physical quantities under
equibiaxial strain. (a) Phonon spectra under strain of 2\% (solid
black lines) and 3\% (dashed orange lines). The sizes of green/red
and blue circles are drawn proportional to the magnitude of the EPC
$\uplambda_{\textbf{\emph{q}}\nu}$ for strain of 2\% and 3\%,
respectively. The red circles have been reduced 20-fold. (b) Band
structures. The Fermi energy level is set at zero. (c)
$\omega$$\rm{_{log}}$ and $N$(\emph{E}$_{\mathrm{{F}}}$) as well as
(d) $T_{c}$ and $\uplambda$ under strain. The lines in (c) and (d) are only guides to the eye.}%
\label{fig4}%
\end{figure}

We note that the 2D boron sheets have been successfully grown on the
Ag(111) substrate \cite{Mannix,Feng}. Growing $\beta_{0}$-PC on
different substrates may also be realized in experiments. In one
recent calculation \cite{Rajbanshi}, strain was found can modulate
the electronic structure and phonon spectrum of $\beta_{0}$-PC.
Here, we want to know the strain effects on the superconductivity,
as already known critical for graphene and phosphorene \cite{Si,Ge}.
Under tensile equibiaxial strain of 0\%$\leq\varepsilon\leq$3\%
where $\varepsilon$=$\frac{a-a_{0}}{a_{0}}\times$100\%, the atomic
structures are fully relaxed. Most of our calculated phonon
dispersions are positive. These results agree well with previous
study \cite{Rajbanshi} where the $\beta_{0}$-PC was found
dynamically stable up to 9.1\% strain. We show in Figure 4a the
phonon dispersions together with the magnitude of the EPC
$\uplambda_{\textbf{\emph{q}}\nu}$ under strains of 2\% and 3\%. We
find that the Kohn anomaly point under strain of 2\% (with some
imaginary frequencies) is moved from near the Y point along
$\Gamma$-Y (under strain of 0\% and 1\%) to near the X point along
$\Gamma$-X. The EPC $\uplambda_{\textbf{\emph{q}}\nu}$ at the Kohn
anomaly point is extremely large. Upon further tensile strain up to
3\%, the Kohn anomaly point over the whole BZ disappears; the phonon
frequencies below 300 cm$^{-1}$ are strengthened while that above
300 cm$^{-1}$ are lowered down; the EPC
$\uplambda_{\textbf{\emph{q}}\nu}$ becomes small.

Under strain, the overall band structure (shown in Figure 4b) is
shifted downward and the $N$(\emph{E}$_{\mathrm{{F}}}$) (Figure 4c)
is decreased firstly to $\sim$7.1 States/spin/Ry/cell
($\varepsilon$=2\%) and then rebounded to $\sim$7.3
($\varepsilon$=3\%). Normally, the superconducting related physical
quantities of $\omega$$\rm{_{log}}$, $\uplambda$, and $T_{c}$ should
increase or decrease monotonously along with applied strain
\cite{Si,Cheng}. However, in our case, the $\omega$$\rm{_{log}}$
decreases firstly to $\sim$90 K ($\varepsilon$=1-2\%) and then
increases to 182.8 K at $\varepsilon$=3\% (Figure 4c); the
$\uplambda$ and $T_{c}$ increase firstly and then decrease (Figure
4d). The abnormal behaviors of these physical quantities are tightly
related to the existence of the Kohn anomaly points under strains of
0-2\% where the EPC strengthes are extremely enlarged. Under
$\varepsilon$=3\%, without the Kohn anomaly point, $\beta_{0}$-PC
still exhibits the superconducting properties of $\uplambda$=1.01
and $T_{c}$=13.03 K. So, the superconducting behavior in
$\beta_{0}$-PC is robust.

Charge carrier doping can also modulate the electronic structure and
phonon spectrum of materials \cite{Savini,Si,MarginePRB,Shao,Xi,Fu}.
In our present study, the carrier doping is simulated by using a
jellium model, whereby the excess/defect electronic charge is
compensated by a uniform neutralizing background
\cite{Si,MarginePRB,Shao,Xi,Fu}. Under different concentrations of
the electron or hole doping, both the lattice constants and the
atomic structures are fully relaxed. As shown in Figure 5, the
electron doping (0.1e/cell) can induce disappearance of the Kohn
anomaly point, raise the phonon frequencies below 300 cm$^{-1}$ a
little up and lower down that above 300 cm$^{-1}$. As a result, the
EPC $\uplambda_{\textbf{\emph{q}}\nu}$ become small and the total
$\uplambda$ is reduced to 0.96. Through electron doping, the overall
band structure is shifted downward and the
$N$(\emph{E}$_{\mathrm{{F}}}$) is increased to $\sim$7.5
States/spin/Ry/cell, the $\omega$$\rm{_{log}}$ is increased to 175
K, and the $T_{c}$ is decreased to 11.56 K. Thus, the electron
doping can suppress the superconducting transition temperature of
$\beta_{0}$-PC.

Under the condition of the hole doping, as indicated in Figure 5,
the Kohn anomaly point near the Y point along $\Gamma$-Y is still
evident. The phonon frequencies below (above) 300 cm$^{-1}$ are
lowered down (raised up) a little. As a result, the EPC
$\uplambda_{\textbf{\emph{q}}\nu}$ on the first acoustic branch are
still large and the total $\uplambda$ are only reduced to 1.43 and
1.07 at doping levels of 0.1 and 0.2 h/cell, respectively. The
overall band structures are shifted upward and the
$N$(\emph{E}$_{\mathrm{{F}}}$) are increased to $\sim$7.6 and 8.0
States/spin/Ry/cell, the $\omega$$\rm{_{log}}$ are increased to 120
and 174 K, and the $T_{c}$ are modulated to 13.09 and 13.47 K. These
results indicate that the hole doping, although changes the
electronic structure and phonon spectrum of $\beta_{0}$-PC to some
extent, induces limited effects on the superconducting transition
temperature.

\begin{figure}
\begin{center}
\includegraphics[width=1\linewidth]{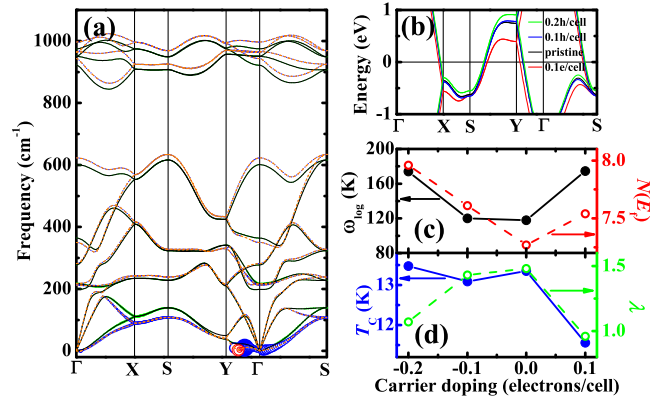}
\end{center}
\caption{Superconducting related physical quantities under carrier
doping. (a) Phonon spectra under electron doping of 0.1 e/cell
(solid black lines) and hole doping of 0.2 h/cell (dashed orange
lines). The sizes of green and blue/red circles are drawn
proportional to the magnitude of the EPC
$\uplambda_{\textbf{\emph{q}}\nu}$ for carrier doping of 0.1 e/cell
and 0.2 h/cell, respectively. The red circles have been reduced
20-fold. (b) Band structures. The Fermi energy level is set at zero.
(c) $\omega$$\rm{_{log}}$ and $N$(\emph{E}$_{\mathrm{{F}}}$) as well
as
(d) $T_{c}$ and $\uplambda$ under carrier doping. The lines in (c) and (d) are only guides to the eye.}%
\label{fig5}%
\end{figure}

In summary, we have investigated the structure, electronic
structure, phonon spectrum, EPC, and superconducting properties of
the phosphorus carbide monolayer ($\beta_{0}$-PC) using
first-principles calculations. We predicted that this intrinsic
monolayer material, the first example within the 2D carbon and
phosphorus families, is an intermediate conventional superconductor.
The out-of-plane P$_{z}$ vibrations together with the lone pair
electrons of P-$p_{z}$ play dominant role for the EPC. The
superconducting behavior in $\beta_{0}$-PC is robust, even under
conditions of tensile equibiaxial strain or electron doping the Kohn
anomaly point has been suppressed. Our findings provide a new choice
in realizing superconductor in 2D limit and will inspire further
efforts in this field.
\section{Acknowledgments}
This work was supported by National Natural Science Foundation of
China under Grant Nos. 11675255 and 11374197.

\end{document}